\documentclass[9pt,twocolumn,twoside]{osajnl}

\journal{ol} 
\usepackage{bm}

\setboolean{shortarticle}{true} 

\title{Electric current induced unidirectional propagation of surface plasmon-polaritons}

\author[1,2,*]{K. Y. Bliokh}
\author[3]{F. J. Rodr\'{i}guez-Fortu\~{n}o}
\author[4]{A. Y. Bekshaev}
\author[2]{Y. S. Kivshar}
\author[1,5]{F. Nori}

\affil[1]{Center for Emergent Matter Science, RIKEN, Wako-shi, Saitama 351-0198, Japan}
\affil[2]{Nonlinear Physics Centre, The Australian National University, Canberra 0200, ACT, Australia}
\affil[3]{Department of Physics, King’s College London, London WC2R 2LC, UK}
\affil[4]{I. I. Mechnikov National University, Dvorianska 2, Odessa, 65082, Ukraine}
\affil[5]{Physics Department, University of Michigan, Ann Arbor, Michigan 48109-1040, USA}

\affil[*]{Corresponding author: kostiantyn.bliokh@riken.jp}

\ociscodes{(240.6680) Surface plasmons; (250.5403) Plasmonics; (230.3810) Magneto-optic systems}

\begin{abstract}
Nonreciprocity and one-way propagation of optical signals is crucial for modern nanophotonic technology, and is typically achieved using magneto-optical effects requiring large magnetic biases. Here we suggest a fundamentally novel approach to achieve unidirectional propagation of surface plasmon-polaritons (SPPs) at metal-dielectric interfaces. We employ a direct electric current in metals, which produces a Doppler frequency shift of SPPs due to the uniform drift of electrons. This tilts the SPP dispersion, enabling one-way propagation, as well as zero and negative group velocities. The results are demonstrated for planar interfaces and cylindrical nanowire waveguides.
\end{abstract}

\setboolean{displaycopyright}{true}

\begin{document}

\maketitle

\vspace{0.5cm}

Nonreciprocity and unidirectional propagation of electromagnetic waves are highly important topics in modern optics, crucial for nanophotonic, quantum-optical, and optoelectronic applications \cite{Potton,Soljacic1,Soljacic2,Fan,Lira,Alu,Khanikaev,Ozdemir,Steinberg,Arno}. 
The main mechanisms generating one-way propagation and strong nonreciprocity are: magneto-optical phenomena \cite{Soljacic1,Soljacic2,Steinberg,Arno,Zvezdin}, including  topological quantum-Hall effect \cite{Soljacic1,Soljacic2}, nonlinearity resulting in optical diodes and circulators \cite{Fan,Ozdemir,Scalora,Tocci,Shadrivov}, and other methods breaking time-reversal symmetry in the system \cite{Lira,Alu}. 

The study of surface waves and plasmonics is another inherent part of nanophotonics, which allows to reduce the length-scales and dimensionality of various electromagnetic phenomena \cite{Zayats,Maier}. Not surprisingly, nonreciprocity and unidirectional propagation of surface plasmon-polaritons (SPPs) have recently attracted considerable attention \cite{Camley,Yu2008,Hu,DE2013,DE2014,Maksymov}. These studies mostly deal with magnetooptical nonreciprocity in the transverse Voigt geometry, including topological quantum-Hall-effect states \cite{DE2013,Silver2015,Silver2016}. 

Here we put forward a novel mechanism resulting in one-way propagation of SPPs at metal-dielectric interfaces. Namely, we show that in the presence of a longitudinal direct electric current, the SPP spectrum becomes nonreciprocal, with a unidirectional propagation in a certain frequency range. This is caused by the Doppler shift of the wave frequency in the drifting electron plasma. Furthermore, the SPP spectrum is deformed such that the group velocity of SPPs propagating along the current vanishes at a critical wavevector, and then becomes negative for larger wavevectors. Thus, the electric-current-induced nonreciprocity is qualitatively different compared to the known magnetic-field-induced case. 

Importantly, we show that the nonreciprocal effect from the electric current can be comparable with the magnetooptical one at reasonable values of the system parameters.
Moreover, we consider SPPs at a planar metal-dielectric interface, as well as in a cylindrical nanowire. Metallic nanowires provide a highly efficient platform for plasmonics and metamaterials \cite{Nanowire1,Nanowire2,Nanowire3,Nanowire4}, and they can be naturally biased by a direct electric current. As we show below, this results in the nonreciprocal properties of nanowire plasmons. 

\begin{figure}[t]
\centering
\includegraphics[width=0.8\linewidth]{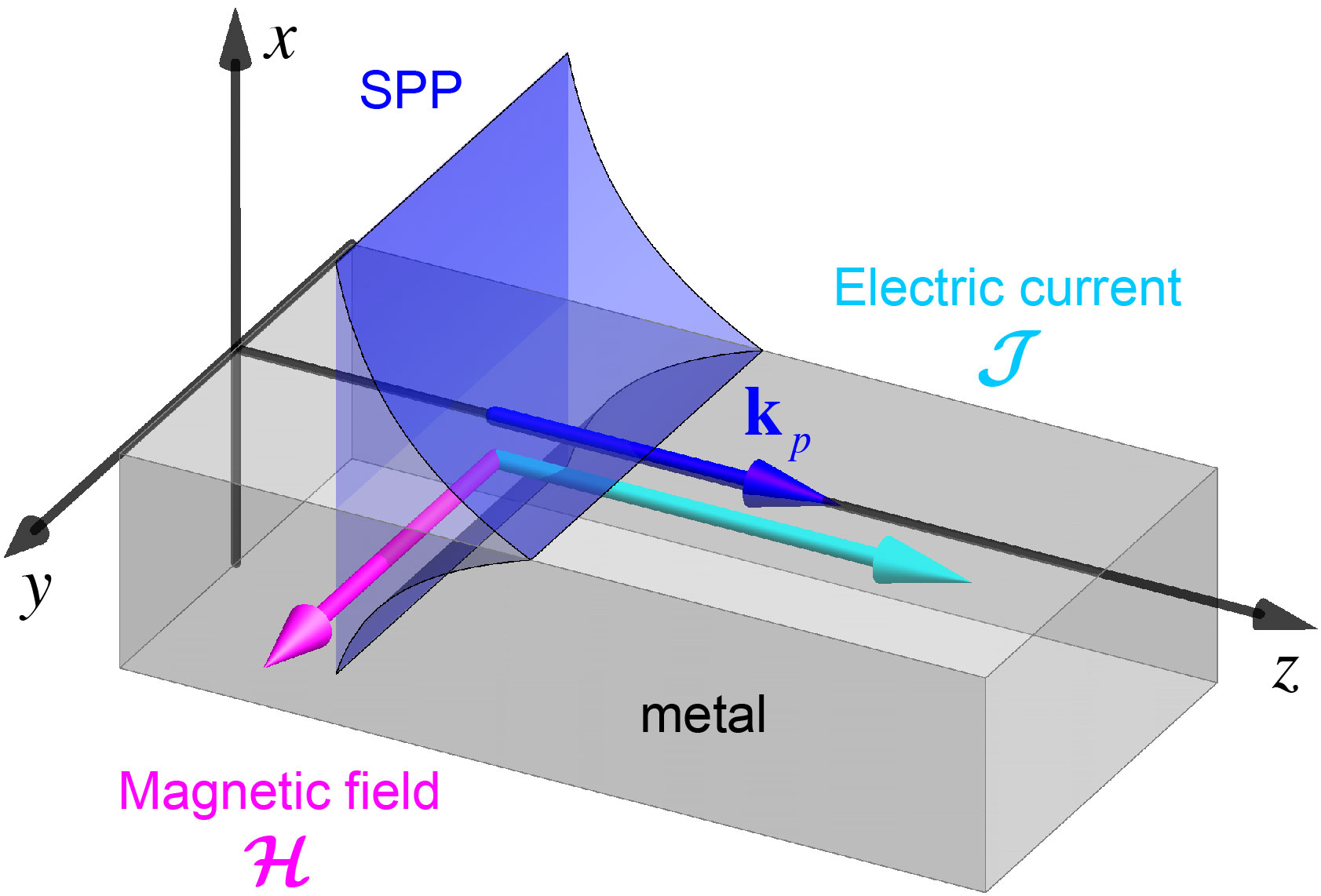}
\caption{Schematic diagram of a surface plasmon-polariton (SPP) propagating along the metal-vacuum interface \cite{Zayats,Maier}. The applied magnetic field ${\bm{\mathcal H}}$, and direct electric current ${\bm{\mathcal J}}$ are shown.}
\label{F1}
\end{figure}

\begin{figure}[t]
\centering
\includegraphics[width=0.8\linewidth]{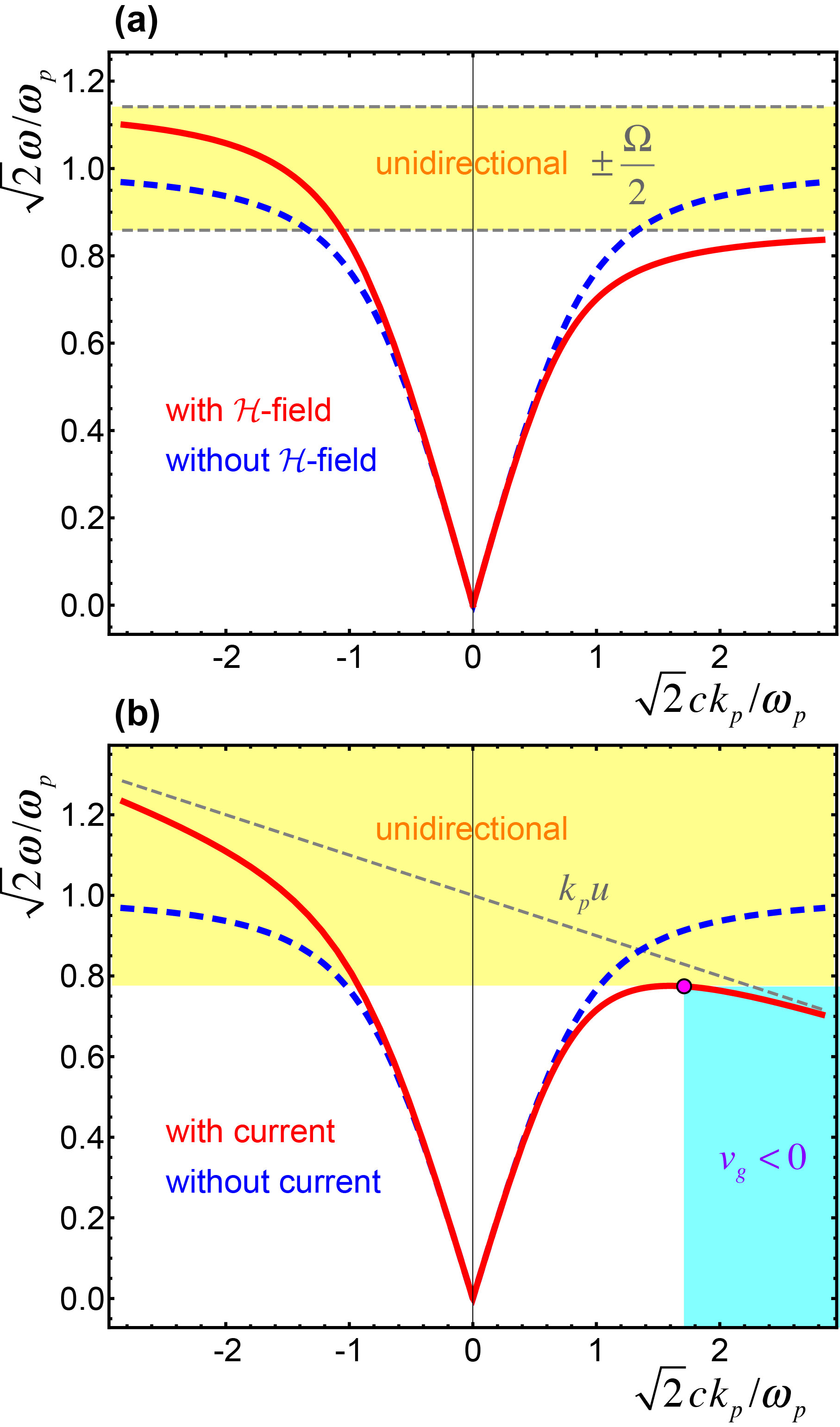}
\caption{Nonreciprocal modifications of the SPP spectra caused by a transverse magnetic field (a), Eq. (\ref{E3}), and a longitudinal direct electric current (b), Eqs. (\ref{E5}) and (\ref{E6}). The dashed curves show the unperturbed reciprocal SPP dispersion ${\omega _0}\left( {{k_p}} \right)$. The frequency ranges with the one-way SPP propagation and the wavevector range with the negative group velocity of SPPs are marked by yellow and blue, respectively. The parameters are $\Omega = 0.2\, \omega_p$ (a) and $u=-0.1\,c$ (b).}
\label{F2}
\end{figure}

To start with, we consider SPPs propagating along the planar metal-vacuum interface $x=0$, in the $\pm z$ directions, as shown in Fig.~\ref{F1}. We employ the simplest Drude model of the metal (neglecting losses) with the permittivity 
$\varepsilon \left( \omega  \right) = 1 - \omega _p^2/{\omega ^2}$ and plasma frequency $\omega_p$. It is well known \cite{Zayats,Maier} that SPPs exist at frequencies $\omega  < {\omega _p}/\sqrt 2$, i.e., $\varepsilon < -1$, and propagate along the interface with wavevector ${{\bf{k}}_p} = {k_p}\,{\bf{\bar z}}$, ${k_p} = \sigma\, {k_0}\sqrt { - \varepsilon } /\sqrt { - 1 - \varepsilon }$, $\left| {{k_p}} \right| > {k_0}$. Hereafter, the overbar denotes the unit vectors of the corresponding axes, ${k_0} \equiv \omega /c$, and we introduced the parameter $\sigma  = {\rm sgn}\, {k_p} =  \pm 1$ indicating the SPP propagation direction. The function ${k_p}\!\left( \omega  \right)$ determines the dispersion relation of SPPs (see the dashed curve in Fig.~\ref{F2}). The SPP field decays away from the interface with the exponential-decay rates ${\kappa _1} = {k_p}/\sqrt { - \varepsilon }$ (in the vacuum, $x>0$) and ${\kappa _2} = \sqrt { - \varepsilon }\, {k_p}$ (in the metal, $x<0$).

We first briefly describe the nonreciprocity and unidirectional propagation of SPPs in the presence of a transverse magnetic field ${{\bm{\mathcal H}}} = {\mathcal H}\,{\bf{\bar y}}$ \cite{Yu2008,Hu,DE2013}. Usually, it is calculated using the anisotropic permittivity tensor of the magnetoactive metal. However, we employ a simpler way to derive the same results. Recently, some of us have shown \cite{Bliokh2017PRL,Bliokh2017NJP} that the $(x,z)$-plane rotation of the electric field of the SPP induces the corresponding orbital motion of electrons in the metal and, hence, the {\it transverse magnetization} due to the inverse Faraday effect. (This property is related to the transverse spin of SPPs \cite{Bliokh2012}, which is currently attracting considerable attention \cite{Bliokh2015PR,Aiello,Bliokh2015NP}.) Using Gaussian units, the magnetization of the metal can be written as \cite{Bliokh2017PRL,Bliokh2017NJP}
%
\begin{eqnarray}
{\bf{M}} = \frac{{ge \omega }}{{4mc}}\, \frac{{d\varepsilon }}{{d\omega }}\, {\rm Im}\! \left( {{{\bf{E}}^*} \times {\bf{E}}} \right) 
= \sigma \, g \, \frac{{\left| E_0 \right|^2}}{\sqrt{-\varepsilon}}\, \frac{{ - e}}{{mc}}\, \exp\!\left( {2{\kappa _2}x} \right){\bf{\bar y}}.
\label{E1}
\end{eqnarray}
%
Here, $g = {\left( {8\pi \omega } \right)^{ - 1}}$, ${\bf{E}}\left( {\bf{r}} \right)$ is the complex electric field in the SPP wave, omitting $\exp\! \left( { - i\omega t} \right)$, $E_0$ is its amplitude right above the metal, whereas $e<0$ and $m$ are the electron charge and mass, respectively. The magnetization (\ref{E1}) means that SPPs, being mixed light-electron quasiparticles, carry {\it transverse magnetic moment} $\bm{\mu}  \propto \sigma \, {\bf{\bar y}}$. It can be calculated as a ratio of the integral magnetization (\ref{E1}) to the number of the quasiparticles. Using the standard Brillouin energy density $W$, this yields \cite{Bliokh2017PRL,Bliokh2017NJP}:
%
\begin{equation}
\bm{\mu}  = \frac{{\hbar \omega }}{{\left\langle W \right\rangle }}\left\langle {\bf{M}} \right\rangle  = \sigma\, \frac{{2\sqrt { - \varepsilon } }}{{1 + {\varepsilon ^2}}}\,{\kern 1pt} {\mu _B}\,{\bf{\bar y}},
\label{E2}
\end{equation}
%
where $\left\langle {...} \right\rangle  = \int {...} \;dx$, and ${\mu _B} = \hbar \left| e \right|/2mc$ is the Bohr magneton. The absolute value of the magnetic moment (\ref{E2}) grows from 0 to $\mu_B$ as the SPP frequency $\omega$ changes from 0 to $\omega_p/\sqrt{2}$.

Equations (\ref{E1}) and (\ref{E2}) describe intrinsic properties of SPPs {\it without} an external magnetic field. Applying the magnetic field ${{\bm{\mathcal H}}} = {\mathcal H}\,{\bf{\bar y}}$ leads to the {\it Zeeman interaction} with the magnetic moment (\ref{E2}), $- \bm{\mu}  \cdot {{\bm{\mathcal H}}}$, which shifts the energy (frequency) of the SPP \cite{QM}. Denoting the SPP frequency without magnetic field as ${\omega _0}\!\left( {{k_p}} \right)$, the Zeeman-shifted frequency in an external magnetic field becomes $\omega\! \left( {{k_p}} \right) = {\omega _0}\!\left( {{k_p}} \right) + \delta \omega\! \left( {{k_p}} \right)$:
%
\begin{equation}
\delta \omega  =  - {\hbar ^{ - 1}} \bm{\mu}  \cdot {{\bm{\mathcal H}}} =  - \frac{{\sqrt { - \varepsilon } }}{{1 + {\varepsilon ^2}}}\, \sigma\, \Omega .
\label{E3}
\end{equation}
%
Here, $\Omega  =  - e {\mathcal H}/m c$ is the cyclotron frequency of the electrons in the magnetic field ${\mathcal H}$, and the correction $\delta\omega$ depends on $k_p$ via $\varepsilon \left[ {{\omega _0}\!\left( {{k_p}} \right)} \right]$. The modified SPP dispersion (\ref{E3}) is shown in Fig.~\ref{F2}(a). The magnetic correction makes the spectrum {\it nonreciprocal}, i.e., depending on the propagation direction $\sigma$. In particularly, the cutoff frequency ${\omega _p}/\sqrt 2$ is now shifted to ${\omega _p}/\sqrt 2  + \sigma \, \Omega /2$. This means that in the range $\omega  \in \left( {{\omega _p}/\sqrt 2  - \Omega /2,{\omega _p}/\sqrt 2  + \Omega /2} \right)$, SPPs become {\it unidirectional}, i.e., propagating only in the positive (negative) $z$-direction for ${\mathcal H} >0$ (${\mathcal H} <0$). Notably, the magnetic correction to the dispersion (\ref{E3}) exactly coincides with the one calculated in \cite{Yu2008} using anisotropic permittivity of the metal in a magnetic field.

We are now in the position to consider SPPs in the presence of a direct electric current with density ${{\bm{\mathcal J}}} = {\mathcal J}\,{\bf{\bar z}}$ flowing in the metal. In this case, the problem can be readily analyzed in terms of the modified permittivity $\varepsilon \left( \omega  \right)$. Indeed, the presence of the current means that free electrons in the metal move with the velocity ${\bf{u}} = {{\bm{\mathcal J}}}/n e$, where $n=m\omega_p^2 / 4\pi e^2$ is the volume density of the electrons. This movement of the electron plasma produces the {\it Doppler frequency shift} $\omega  \to \omega  - k\, u$ in the metal permittivity \cite{PK}:
%
\begin{equation}
\varepsilon \left( \omega  \right) = 1 - \frac{{\omega _p^2}}{{{{\left( {\omega  - {k_p}u} \right)}^2}}} .
\label{E4}
\end{equation}
%
Considering the $z$-aligned propagation of SPPs, we can still employ the usual form of the SPP dispersion relation, ${k_p} = {k_0}\,\sqrt { - \varepsilon } /\sqrt { - 1 - \varepsilon }$, but now with the Doppler-modified permittivity (\ref{E4}). Expanding this in the linear approximation in the drift velocity $u$, we arrive at the following dispersion relation:
%
\begin{equation}
\omega  = {\omega _0}\!\left( {{k_p}} \right) + \frac{{1 - \varepsilon }}{{1 + {\varepsilon ^2}}}\,\sigma \left| {{k_p}} \right| u .
\label{E5}
\end{equation}
%

The current-modified SPP dispersion (\ref{E5}) is nonreciprocal, as shown in Fig.~\ref{F2}(b). Moreover, this nonreciprocity differs qualitatively from the known magnetic-field case, Fig.~\ref{F2}(a). Indeed, the cut-off frequency asymptote ${\omega _p}/\sqrt 2$ (for $\left| {{k_p}} \right| \to \infty$) is now {\it tilted} as ${\omega _p}/\sqrt 2  + {{k_p}} u$, rather than split. The most interesting feature of the modified dispersion is that it has an {\it inflexion point}:
%
\begin{equation}
k_p^{\rm inf} =  - \frac{{{\omega _p}}}{{\sqrt{2}\,c}}{\left( {\frac{c}{2\,u}} \right)^{\frac{1}{3}}} , \quad
\omega\! \left( {k_p^{\rm inf}} \right) = \frac{{{\omega _p}}}{{\sqrt 2 }} + \frac{3}{2}\,k_p^{\rm inf}\,u .
\label{E6}
\end{equation}
%
The SPP group velocity ${v_g} = \partial \omega /\partial {k_p}$ vanishes and changes its sign in this point. For positive current ${\mathcal J} >0$, $u<0$, $k_p^{\rm inf} > 0$, and the group velocity becomes negative for $k_p >k_p^{\rm inf}$. This is because slow SPPs near the cut-off frequency ${\omega _p}/\sqrt 2$ are dragged by the flow of electrons in the backward direction. Furthermore, the inflexion point (\ref{E6}) determines the maximum frequency of the SPPs propagating along the current ${\bm{\mathcal J}}$. For $\omega > \omega\! \left( {k_p^{\rm inf }} \right)$, SPPs become {\it unidirectional}, propagating only in the direction opposite to the current. Due to the tilt of the cut-off asymptote, the unidirectional-propagation range is not limited from above by a higher frequency. However, practically, high wave numbers $\left| {{k_p}} \right|$ are accompanied by strong absorption of SPPs \cite{Maier}. 

The inflexion-point parameters are determined by the ratio of the electron drift velocity to the speed of light: $\left| u \right|/c \ll 1$. For typical laboratory currents, this is a very small parameter. However, the power of 1/3 makes the inflexion-point characteristics not too extreme, resulting in observable consequences at feasible parameters. In particular, the current-induced cut-off frequency shift $\sim k_p^{\rm inf }u$ could be of the order of or even larger than the similar magnetic-field-induced shift $\sim \Omega$. 

\begin{figure}[t]
\centering
\includegraphics[width=0.8\linewidth]{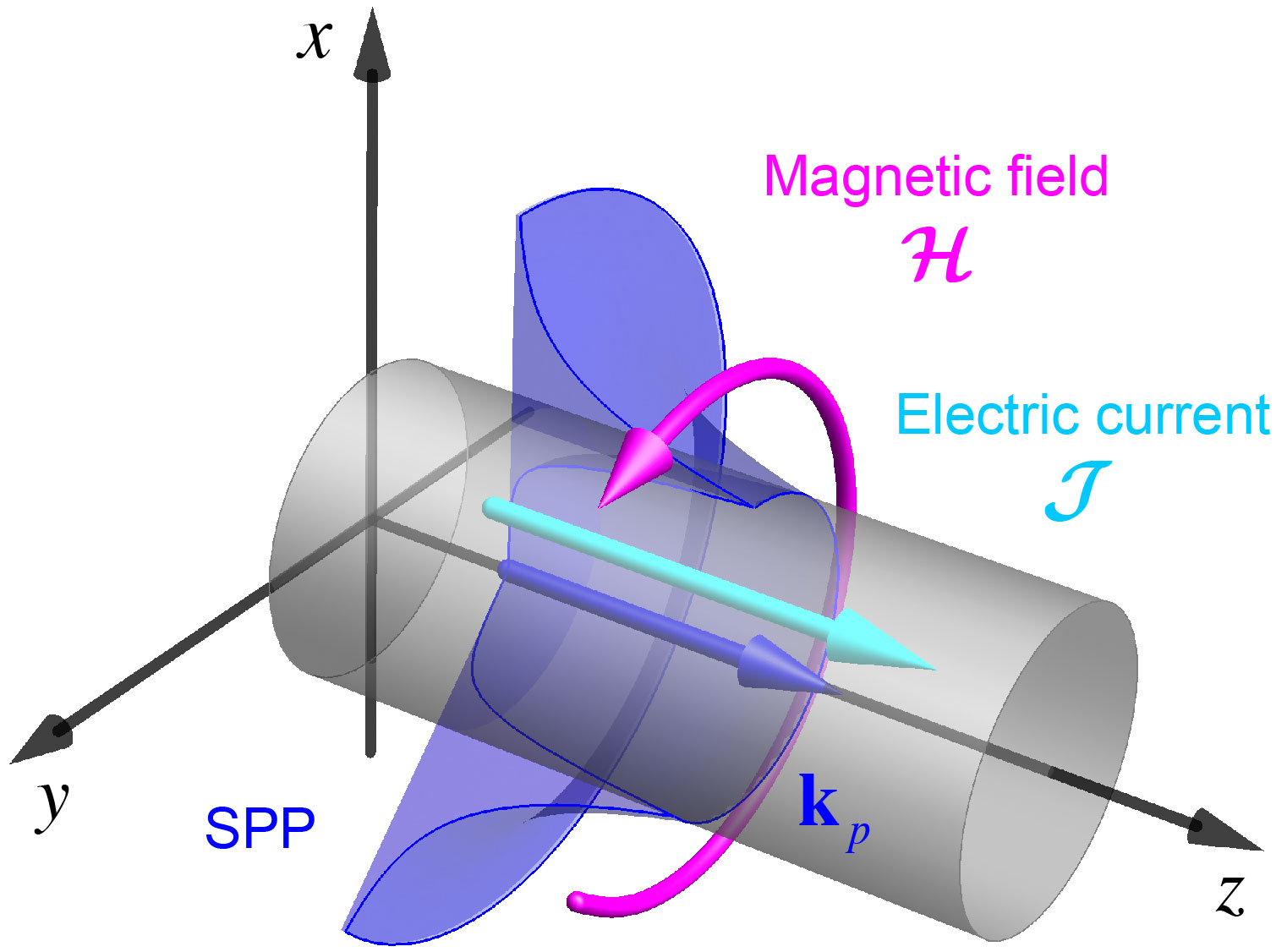}
\caption{Schematic diagram of a SPP mode of a metallic nanowire. The direct electric current ${\bm{\mathcal J}}$ in the nanowire and the corresponding induced magnetic field ${\bm{\mathcal H}}$ (considered in \cite{DE2014}) are shown.}
\label{F3}
\end{figure}

For example, the work \cite{DE2014} considered a gold nanowire of radius $r_0=10^{-5}\,$cm in the presence of an electric current ${\mathcal I} = \pi\, r_0^2\,{\mathcal J} = 75 \cdot {10^{ - 3}}\,$A, see Fig.~\ref{F3}. Using the free-electron density in gold, $n \simeq 6 \cdot {10^{22}}\,{\rm cm}^{-3}$, we find the electron drift velocity $\left| u \right| \simeq 2.5 \cdot {10^4}\,{\rm cm}/{\rm s} \simeq 0.8 \cdot {10^{ - 6}}\, c$. The SPP cut-off frequency was ${\omega _p}/\sqrt 2  \simeq 4.8 \cdot {10^{15}}\, {\rm s}^{-1}$, and the SPP wave number $\left| {{k_p}} \right| \simeq 2 \cdot {10^7}\,{\rm cm}^{-1}$. The work \cite{DE2014} examined the nonreciprocal effect of the azimuthal magnetic field generated by the current, ${{\mathcal H}_{\varphi }} = 2{\mathcal I}/c{r_0} \simeq 1.5 \cdot {10^3}\,$G (and enhanced by a magnetoactive dielectric around the wire), but neglected the direct electric-current effect on surface plasmons. In fact, the above parameters correspond to the cyclotron frequency $\Omega  \simeq 3 \cdot {10^{10}}\, {\rm s}^{-1}$ and the Doppler frequency shift $\left| {{k_p}u} \right| \simeq 5 \cdot {10^{11}}\, {\rm s}^{ - 1} \gg \Omega$.   Moreover, the chosen wavenumber exactly corresponds to the inflexion point (\ref{E6}): $| k_p | \simeq |k_p^{\rm inf}|$, where the group velocity in the current direction vanishes and only the backward propagation is possible. Thus, the electric-current nonreciprocity is stronger than the magnetic-field one (in a pure metal, without a magnetoactive dielectric), and it can provide one-way propagation for these parameters.

To properly analyze the electric-current effect in a nanowire, we now consider SPPs in the cylindrical geometry of a metallic wire of radius $r_0$, Fig.~\ref{F3}. For the sake of generality, we introduce the permittivity $\varepsilon_1 >0$ and permeability $\mu_1 >0$ outside the wire and the permittivity $\varepsilon_2 <0$ and permeability $\mu_2 >0$ inside the wire (later we set ${\varepsilon_1} = {\mu_1} = {\mu_2} = 1$). The fundamental plasmonic wire mode is TM polarized and, hence, can be described by the vector potential ${\bf{A}} = A\,{\bf{\bar z}}$ \cite{Balanis}, where $A$ is the zero-order solution of the scalar wave equation in cylindrical coordinates $\left( {r,\varphi ,z} \right)$:
%
\begin{equation}
A = A_0\, {e^{i{k_p}z}}\left\{ {\begin{array}{*{20}{c}}
{a\,{I_0 }\!\left( {{\kappa _2}r} \right),\quad r < {r_0}}\, , \\
{b\,{K_0 }\!\left( {{\kappa _1}r} \right),\quad r > {r_0}}\, .
\end{array}} \right.
\label{E7}
\end{equation}
%
Here, $k_p$ is the propagation constant, ${\kappa _{1,2}} = \sqrt {k_p^2 - k_{1,2}^2}$ are the radial exponential-decay constants, ${k_{1,2}} = \sqrt {{\varepsilon _{1,2}}{\mu _{1,2}}} \,{k_0}$ are the wave numbers in the two media, while $I_0$ and $K_0$ are the modified Bessel functions. The amplitudes  $(a,b)$ in Eq.~(\ref{E7}) are to be determined. The wave electric and magnetic fields in each medium are given by \cite{Balanis}:
%
\begin{eqnarray}
{\bf{E}} = i{k_0}{\bf{A}} + \frac{{i {k_0}}}{{k_{1,2}^2}}\bm{\nabla} \left( {\bm{\nabla}  \cdot {\bf{A}}} \right) , \quad
{\bf{H}} = \frac{1}{{{\mu _{1,2}}}}\bm{\nabla}  \times {\bf{A}}\, .
\label{E8}
\end{eqnarray}
%
Substituting the potential (\ref{E7}) into Eqs. (\ref{E8}), we obtain all vector components of the wave fields. Applying the electromagnetic boundary conditions at $r=r_0$, we arrive at the system of equations for the amplitudes $(a,b)$:
%
\begin{equation}
\hat M\! \left( {\begin{array}{*{20}{c}}
a\\
b
\end{array}} \right) \! \equiv \! \left( {\begin{array}{*{20}{c}}
{\kappa _2^2{\varepsilon _1}{\mu _1}{I_0}\!\left( {{\rho _2}} \right)}&{ - \kappa _1^2{\varepsilon _2}{\mu _2}{K_0}\! \left( {{\rho _1}} \right)}\\
{2{\kappa _2}{\mu _1}{I_1}\! \left( {{\rho _2}} \right)}&{2{\kappa _1}{\mu _2}{K_1}\! \left( {{\rho _1}} \right)}
\end{array}} \right)\! \left( {\begin{array}{*{20}{c}}
a\\
b
\end{array}} \right) \!=\! 0\, ,
\label{E9}
\end{equation}
%
where ${\rho _{1,2}} = {\kappa _{1,2}}{r_0}$. Equation (\ref{E9}) has non-trivial solutions only when $D \equiv {\rm det}\, \hat M = 0$, which provides the transcendental characteristic equation $D\! \left( {\omega ,{k_p}} \right) = 0$ for the plasmonic mode dispersion. 

\begin{figure}[t]
\centering
\includegraphics[width=0.8\linewidth]{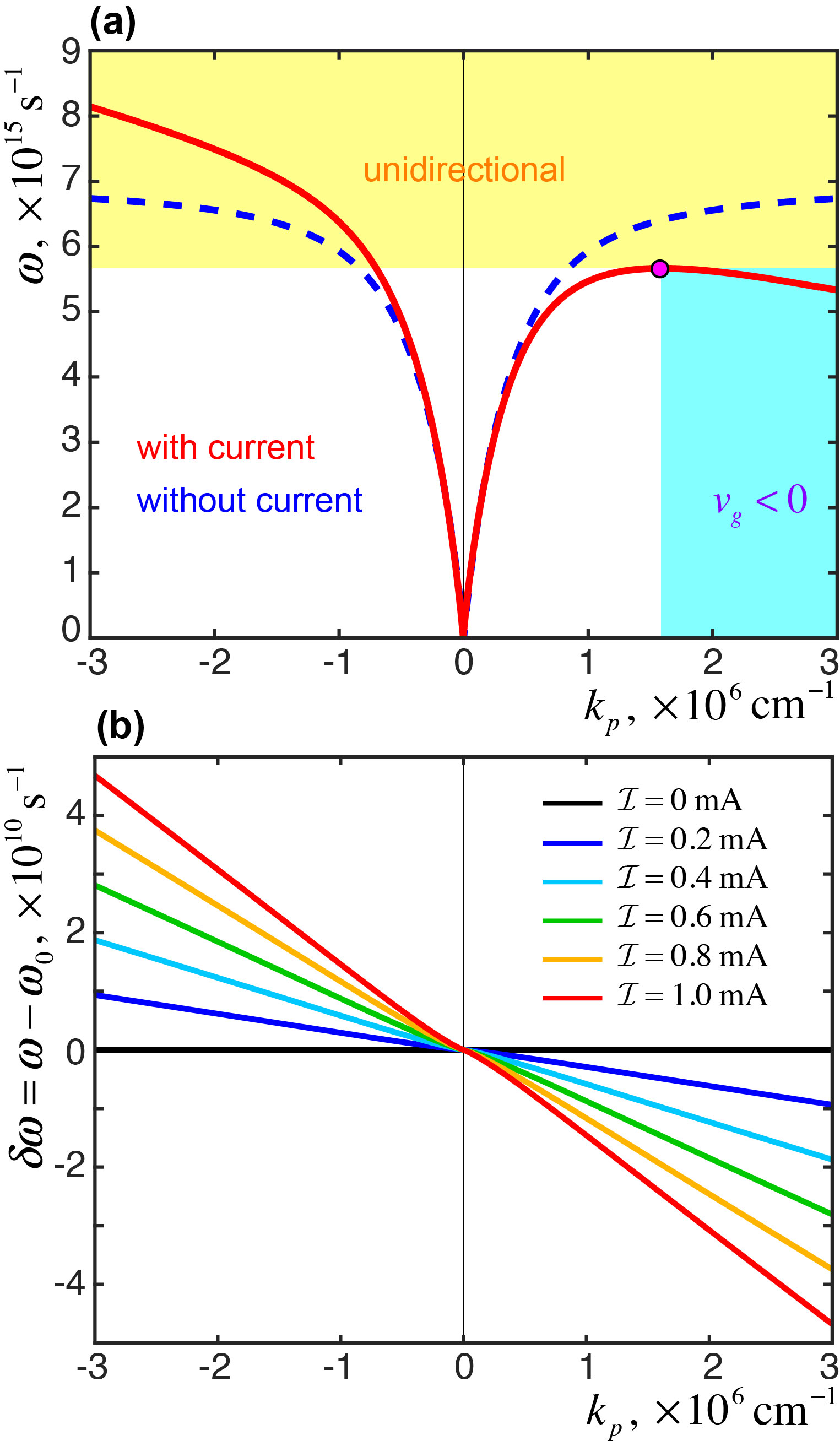}
\caption{Nonreciprocal electric current-induced modifications of the SPP spectra in a metallic nanowire with ${\omega _p} = {10^{16}}\,{\rm s}^{-1}$ and radius ${r_0} = 20\,$nm. (a) The unidirectional-propagation and negative-group-velocity ranges are shown for a very high current ${\mathcal I}=30\,$A, cf. Fig. 2b. (b) Small nonreciprocity from realistic currents ${\mathcal I} \leq 1\,$mA is depicted in the form of the deviation $\delta \omega\! \left( {{k_p}} \right)$ from the reciprocal dispersion $\omega_0\! \left( {{k_p}} \right)$.}
\label{F4}
\end{figure}

Similarly to the planar SPP case, we introduce the effect of the electric current via the Doppler shift (\ref{E4}) in the Drude-metal permittivity ${\varepsilon _2} \equiv \varepsilon \left( \omega  \right)$. The drift velocity of the electrons is related to the current as ${\mathcal I} = \pi r_0^2\,{\mathcal J} = \pi r_0^2\,neu$. Substituting the Doppler-modified permittivity (\ref{E4}) into the characteristic equation, we numerically find the modified dispersion relation for the fundamental SPP mode in the electric-biased nanowire. Figure~\ref{F4} shows the dispersion relation for a nanowire with ${\omega _p} = {10^{16}}\,{\rm s}^{-1}$, ${r_0} = 20\,$nm, and different values of ${\mathcal I}$. Panel (a) shows the modified dispersion for a very high value of the current ${\mathcal I}=30\,$A, chosen to exaggerate the nonreciprocal effect, while panel (b) displays the zoomed-in perturbation of the SPP dispersion for realistic smaller currents ${\mathcal I} \leq 1\,$mA. All the features discussed for the planar SPP, including the one-way propagation and negative group velocity ranges, can be clearly observed here. 

In conclusion, we have proposed a simple but yet fundamental way to achieve unidirectional propagation of surface plasmon-polaritons using a direct electric current in  metals. The one-way propagation of optical signals, in analogy to electronic isolators, is considered as a fundamental requirement for enabling photonic high-speed all-optical processing that could substitute current microelectronic components. Nonreciprocal propagation requires breaking the time-reversal symmetry in the system. This is usually done via magneto-optical effects requiring large magnetic biases. In contrast, our proposal is based on the use of an electric current, which can be naturally generated in plasmonic waveguides. The ability to achieve one-way optical propagation using direct electric currents is conceptually simple and inherently compatible with modern microelectronics industry. This approach does not require bulky external magnets and can be easily implemented in an on-chip integrated environment potentially combining electrical and optical components.

We acknowledge helpful discussions with Y. P. Bliokh, N.~Noginova, M. Durach, and M. F. Picardi. This work was supported by the RIKEN iTHES Project, MURI Center for Dynamic Magneto-Optics via the AFOSR Award No. FA9550-14-1-0040, the Japan Society for the Promotion of Science (KAKENHI), the IMPACT program of JST, CREST grant No. JPMJCR1676, the John Templeton Foundation, European Research Council (Starting Grant ERC-2016-STG-714151-PSINFONI), and the Australian Research Council.

\bigskip

\bibliography{sample}

\ifthenelse{\equal{\journalref}{aop}}{%
\section*{Author Biographies}
\begingroup
\setlength\intextsep{0pt}
\begin{minipage}[t][6.3cm][t]{1.0\textwidth} 
  \begin{wrapfigure}{L}{0.25\textwidth}
    \includegraphics[width=0.25\textwidth]{john_smith.eps}
  \end{wrapfigure}
  \noindent
  {\bfseries John Smith} received his BSc (Mathematics) in 2000 from The University of Maryland. His research interests include lasers and optics.
\end{minipage}
\begin{minipage}{1.0\textwidth}
  \begin{wrapfigure}{L}{0.25\textwidth}
    \includegraphics[width=0.25\textwidth]{alice_smith.eps}
  \end{wrapfigure}
  \noindent
  {\bfseries Alice Smith} also received her BSc (Mathematics) in 2000 from The University of Maryland. Her research interests also include lasers and optics.
\end{minipage}
\endgroup
}{}

\end{document}